\documentclass[12pt]{iopart}
\usepackage{iopams}
\usepackage{graphicx}
\usepackage{amssymb}
\newcommand{\gguide}{{\it Preparing graphics for IOP Publishing journals}}
\usepackage{graphicx}
\usepackage{dcolumn}
\usepackage{bm}
\usepackage[colorlinks=true,citecolor=blue,linkcolor=blue,urlcolor=blue]{hyperref}

\usepackage{tikz}
\usepackage{soul}
\usepackage{amsmath}

\usepackage{pstricks}
\usepackage{color}
\usepackage{slashed}
\usepackage{epsfig}
\usepackage{amsfonts}
\usepackage[giveninits=true, sorting=none, backend=biber, style=numeric-comp, maxnames=10]{biblatex}
\bibliography{main.bib}
\def\beq{\begin{equation}}
\def\eeq{\end{equation}}
\def\bea{\begin{eqnarray}}
\def\eea{\end{eqnarray}}
\def\nn{\nonumber}
\def\lan{\langle}
\def\ran{\rangle}
\def\r{\mathbf{r}}
\def\Re{\textrm{Re}}
\def\Im{\textrm{Im}}
\def\Tr{\textrm{Tr}}
\def\x{\mathbf{x}}
\def\y{\mathbf{y}}
\def\z{\mathbf{z}}
\def\p{\mathbf{p}}
\def\d{\mathrm{d}}
\def\sbullet{{\scriptstyle \bullet}}
\def\q_perp{\mathbf{q}_{\perp}}
\newcommand{\gsm}[1]{\textcolor{red}{#1}}

\newcommand{\bfs}[1]{\textcolor{blue}{#1}}

\RequirePackage[normalem]{ulem}
\newcommand{\new}[1]{{\textcolor{blue}{#1}}}
\newcommand{\old}[1]{\textcolor{red}{\sout{#1}}}
\newcommand{\brown}[1]{{\textcolor{brown}{#1}}}

\newcommand{\inner}[2]{\langle #1,#2\rangle} 
\newcommand{\Inner}[2]{\left\langle #1,#2\right\rangle}
\newcommand{\norm}[1]{\|{#1}\|}
\newcommand{\Norm}[1]{\left\|{#1}\right\|}
\newcommand{\abs}[1]{|#1|}
\newcommand{\Abs}[1]{\left|#1\right|}
\newcommand{\R}{\mathbb{R}}

\begin{document}

\title{
Weak- and strong-disorder effects in the continuous random-field Ising model}
\author{G. O. Heymans$^1$\footnote{Author to whom any correspondence should be addressed.}, N.~F.~Svaiter$^1$, B.~F.~Svaiter$^2$, A. M. S. Macêdo$^3$}

\address{$^1$Centro Brasileiro de Pesquisas F\'{\i}sicas - CBPF, \\ Rua Dr. Xavier Sigaud 150 22290-180 Rio de Janeiro, RJ, Brazil}

\address{$^2$Instituto de Matemática Pura e Aplicada - IMPA \\ Estrada Dona Castorina 110 22460-320 Rio de Janeiro, RJ, Brazil}

\address{$^3$ Universidade Federal de Pernambuco - UFPE \\ Av. Prof. Moraes Rego 1235 50670-901, Recife, PE, Brazil}

 \eads{\mailto{olegario@cbpf.br},
\mailto{nfuxsvai@cbpf.br},
\mailto{benar@impa.br}, \mailto{antonio.smacedo@ufpe.br}} 
\vspace{10pt}
\begin{indented}
\item[]February 2026
\end{indented}


\begin{abstract}
We discuss the weak- and strong-disorder
effects in the continuous random-field Ising model using the distributional zeta-function method. In the weak-disorder regime, the vacuum is located at the origin, whereas in the strong-disorder regime, the vacuum is shifted from the origin. By performing the quenched-disorder average at the level of the
effective action, we derive the quadratic and interaction terms. In the weak-disorder limit, we show that the infrared structure of the two-point correlation functions yields a
decomposition of the physical field into correlated components with distinct scaling
dimensions. This mechanism exhibits the characteristic $1/p^4$ behavior, which shifts
the upper critical dimension to $d_c^{+}=6$. The universal critical behavior of the RFIM
near this dimension is governed by a minimal infrared effective action. In the
strong-disorder regime, we obtain a  diagonal quadratic action with a discrete
spectrum of massive modes. The absence of massless modes implies that conventional criticality may be replaced by a symmetry-restoration transition. The resulting spectral representation of correlation
functions converges rapidly and remains well controlled in the infrared limit.
\end{abstract}


\pacs{05.20.-y, 75.10.Nr}

\maketitle


\section{Introduction}

In the study of disordered systems, 
the random-field Ising model (RFIM) has been the subject of extensive theoretical
\cite{nattermann1989random,young1998spin}, experimental, and numerical studies
\cite{fytas2016phase, Fytas:2016itl}. This model was introduced by Imry and Ma in
1975 \cite{Imry:1975zz} as a magnetic counterpart to the effect predicted by
Larkin in 1970 \cite{larkin1970effect}, namely the destruction of  long-range translational
order in the Abrikosov vortex lattice due to the presence of weak impurities. The RFIM
can be realized in the laboratory \cite{fishman1979random} by applying a uniform external
field to a diluted Ising antiferromagnet. Other well-known experimental realizations of
the RFIM include binary liquids in random porous media \cite{de1984liquid}. The RFIM on
a hypercubic $d$-dimensional lattice is defined by the Hamiltonian
\begin{equation}
H=-J\sum_{(i,j)=1}^{N}\,S_{i}S_{j}-\sum_{i=1}^N\,h_{i}S_{i}, \label{1}
\end{equation}
where $(i,j)$ indicates that the sum is taken over nearest-neighbor pairs, $S_{i}=\pm 1$,
$N$ is the total number of Ising spins, and the $h_{i}$ are quenched random variables.
Periodic boundary conditions are typically imposed, and the thermodynamic limit is taken at the end.
The physical properties of the system can be obtained from its partition function, $Z=\Tr\, e^{-\beta H}$.
The \emph{quenched} free energy, $F$, is defined as
$\beta F=-\mathbb{E}[\ln Z]$, where $\mathbb{E}[\cdots]$ denotes the average over the ensemble of
all disorder realizations, which we take to be normally distributed.
%
%
The probability distribution of these quenched random variables has zero mean,
$\mathbb{E}[h_{i}]=0$, and correlation functions given by $\mathbb{E}[h_{i}h_{j}]=h_{0}^{2}\delta_{ij}$.

A central question in the physics of disordered systems is how the critical behavior of a
system is modified in the presence of disorder \cite{Harris:1974zug}. In particular, what is the
nature of the transition from the symmetric to the ordered phase?
Two dimensions are of particular relevance in pure and quenched disordered models.
The first is the lower critical dimension $d_{c}^{\,-}$, i.e., the smallest spatial dimension
in which long-range order can exist.
The second is the upper critical dimension $d_{c}^{\,+}$, above which the model becomes
Gaussian in the infrared.
In Ref. \cite{Imry:1975zz}, two results were presented for the RFIM in the regime dominated by
disorder fluctuations.
First, using Peierls arguments \cite{peierls1936ising}, the authors showed that $d_{c}^{\,-}=2$.
Second, they showed that $d_{c}^{\,+}=6$ by means of renormalization-group techniques.
The first result was discussed by Imbrie \cite{Imbrie:1984ki}, and the second was confirmed by
Aizenman and Wehr \cite{Aizenman:1989dk, Aizenman:1990ji}.
Concerning the existence of the phase transition, Refs. \cite{Bricmont:1987fv, bricmont1988phase} showed that an ordered phase exists for $d\geq 3$. See also Ref. \cite{grinstein1984lower}.
Further important results comparing the behavior of pure and disordered models were obtained
by many authors. It was shown that the critical exponents of a system with quenched disorder are
identical to those of the corresponding pure system in $(d-2)$ dimensions
\cite{grinstein1976ferromagnetic, Aharony:1976jx, Parisi:1979ka, Kaviraj:2021qii, Cardy:2023zna,klein1984supersymmetry}.
While this dimensional reduction breaks down at low dimensions ($d<5$), recent high-precision
numerical studies \cite{fytas2018review} have demonstrated that it holds with remarkable accuracy
at $d=5$. 

We also acknowledge the complementary perspective provided by functional renormalization-group
methods developed by Tarjus, Tissier, and collaborators \cite{tarjus2020random,dupuis2021nonperturbative}
for studying the RFIM. While methodologically distinct, both approaches focus on similar physical observables: these works analyze moments of the disordered free energy, paralleling our analysis of partition function moments and suggesting potential for future synthesis of these perspectives.
%
%

The intriguing interplay between disorder and criticality in the RFIM---especially the dimensional
reduction property---provides fertile ground for exploring applications beyond its original
formulation. Recent studies of quasi-long-range order in the RFIM reveal how local Markov
statistics can generate intermediate states between long-range order and disorder. This
emergence of complex behavior from simple local rules resonates with phenomena observed in diverse
fields in which Markov random fields are applied, such as image analysis, network science, and
social dynamics \cite{Enrique}.

The soft-spin version of the RFIM is a Landau--Ginzburg model with additive quenched disorder.
The aim of this paper is to study the critical properties of a Landau--Ginzburg model in the
presence of additive quenched disorder, i.e., a nonthermal control parameter.
%
%
%
In this setting, the disorder-averaged free energy is a particularly relevant physical quantity 
 \cite{englert1963linked, griffiths1968random}.
%
One way to obtain this average is through the replica trick
\cite{edwards1975theory, Emery:1975zz,mezard1987spin, dotsenko2005introduction}. %
Other possibilities include dynamical analyses of systems with quenched disorder
\cite{de1978dynamics,sompolinsky1981dynamic,de2006random}, as well as the use of Grassmann
anticommuting variables \cite{Efetov:1983xg}.
An alternative method discussed in the literature is the
distributional zeta-function method (DZF) \cite{Svaiter:2016lha,Svaiter:2016-2, Diaz:2017grg,Diaz:2017ilf, Diaz:2016mto}. In this formalism, the quenched disorder-averaged free energy can be obtained without using an analytic extension procedure.


%
%

In this work, we use the DZF method to analyze the random-field Ising model in both the weak- and strong-disorder regimes. In the weak-disorder regime, the vacuum is located at the origin, whereas in the strong-disorder regime, the vacuum is shifted from the origin. In the weak-disorder limit, we derive the
disorder-averaged effective action and show how its infrared structure naturally leads to the
emergence of correlated scaling fields with distinct dimensions, reproducing the characteristic
$1/p^4$ behavior, the upper critical dimension $d_c^{+}=6$, and the Cardy decomposition. This
allows us to identify the infrared-relevant operators
governing the universal behavior of the RFIM and to construct a minimal effective action that
encodes its critical physics near the upper critical dimension. In the strong-disorder limit, we
obtain a diagonal quadratic action with a discrete spectrum of massive modes; the absence of massless modes at tree-level implies that conventional criticality may be replaced by a symmetry-restoration transition. Taken together, these results shed light on the
mechanisms by which quenched disorder reshapes infrared physics in the RFIM.

Let us briefly review the functional formalism for pure continuous systems, i.e., soft-spin systems without disorder \cite{Wilson:1973jj}.
In such a pure setting, the Ising Hamiltonian is replaced by the Landau--Ginzburg functional
\begin{equation}
S(\varphi)=
\int d^{d}x\left[\frac{1}{2}\varphi(x)\left(-\Delta+m_{0}^{2}\right)\varphi(x)+\frac{\lambda_{0}}{4!}\varphi^{4}(x) \right],
\label{9}
\end{equation}
%
where $\Delta$ denotes the Laplacian in $\mathbb{R}^{d}$, $\lambda_{0}$ is the bare coupling
constant, and $m_{0}$ is a spectral parameter of the model, usually referred to as the mass.
The partition function of the system, $Z$, is defined as a functional integral
\begin{equation}\label{eq:part}
Z=\int_{\partial\Omega} [\d \varphi]\,\,\exp\bigl(-S(\varphi)\bigr),
\end{equation}
where $[\d \varphi]$ is the functional measure, formally given by $[\d \varphi]=\prod_{x} \d\varphi(x)$,
$\partial\Omega$ denotes the boundary of the domain, and its appearance in the functional integral
indicates that the field $\varphi(x)$ satisfies prescribed boundary conditions. To preserve translational
invariance, periodic boundary conditions may be imposed by replacing $\mathbb{R}^{d}$ with the
torus $\mathbb{T}^{d}$. The thermodynamic, or infinite-volume, limit is then taken at the end.

From the partition function, one constructs a probability measure equivalent to the Gibbs measure
of statistical mechanics, and the expectation value of any polynomial functional of the field,
$f(\varphi)$, is given by
\begin{equation}
\langle f(\varphi) \rangle =\frac{1}{Z}\int [d\varphi]f(\varphi) \exp\bigl(-S(\varphi)\bigr).
\end{equation} 
Using these definitions, all $n$-point correlation functions of the model can be computed.
Introducing an external source $j(x)$, one can define the generating functional for all
$n$-point correlation functions, $Z(j)$, as
%
\begin{equation}
Z(j)=\int_{\partial\Omega} [d\varphi]\,\, \exp\left(-S(\varphi)+\int d^{d}x j(x)\varphi(x)\right).
\label{eq:generatingfunctional}
\end{equation}
Next, using the linked-cluster theorem, one defines the generating functional of
connected correlation functions, $W(j)=\ln Z(j)$. Both generating functionals may be represented
by Volterra series.
By taking functional derivatives with respect to the external source and then setting $j(x)=0$, one
obtains the $n$-point and connected $n$-point correlation functions of
the model, respectively.

To treat disordered systems, we introduce the functional $Z(j,h)$, the generating functional of
correlation functions for a fixed disorder realization $h(x)$. As before, $j(x)$ denotes an external
source. This functional integral is defined as
\begin{equation}
Z(j,h)=\int_{\partial\Omega} [d\varphi]\,\, \exp\left(-S(\varphi,h)+\int \d^{d}x\, j(x)\varphi(x)\right),
\label{eq:disorderedgeneratingfunctional}
\end{equation}
where the action functional in the presence of disorder, $S(\varphi, h)$, is given by 
\begin{equation} 
S(\varphi,h)=S(\varphi)+ \int \d^{d}x\,h(x)\varphi(x).
\end{equation}
In the above equation, $S(\varphi)$ is the pure Landau--Ginzburg action functional defined in
Eq.~(\ref{9}), and $h(x)$ is 
a quenched disorder.

To compute the quenched disorder average, one considers $\ln Z(j,h)$ for a fixed
disorder realization $h(x)$ and then averages over all disorder realizations.
As in the pure-system case, one can define the generating functional of connected correlation
functions for a fixed disorder realization as $W(j,h)=\ln Z(j,h)$.
Therefore, we define the disorder average of $W(j,h)$ as the following
generating functional:
\begin{equation}
\mathbb{E}\bigl[W(j,h)\bigr]=\int\,[\d h]P(h)\ln Z(j,h).
\label{eq:disorderedfreeenergy}
\end{equation} 
The conventional way to derive the connected correlation functions is to take functional derivatives of
$\mathbb{E}[W(j,h)]$ with respect to $j(x)$ directly in Eq.~(\ref{eq:disorderedfreeenergy}).
Note that, to compute all correlation functions of the model, one has to deal with the
contribution $(Z(j,h))^{-1}$. Although the theory is averaged over the disorder and one obtains a model without
spatial heterogeneities, the effects of local disorder fluctuations in the original system
still appear in this formalism.

\section{Distributional zeta-function method}

For a general disorder probability distribution, we use the disorder-dependent functional integral
$Z(j,h)$ given by Eq.~(\ref{eq:disorderedgeneratingfunctional}) to define the distributional zeta-function $\Psi(s)$ as
\begin{equation}\label{eq:zetager}
\Psi(s)=\int [dh]P(h)\frac{1}{Z(j,h)^{s}},
\vspace{.2cm}
\end{equation}
\noindent for $s\in \mathbb{C}$, defined in the region where the above integral converges.
The average generating functional can be written as
\begin{equation}
\mathbb{E}\bigl[W(j,h)\bigr]=-(d/ds)\Psi(s)|_{s=0^{+}}, \,\,\,\,\,\,\,\,\,\, \Re(s) \geq 0,  
\end{equation}
where one defines the complex exponential $n^{-s}=\exp(-s\log n)$, with $\log n\in\mathbb{R}$.
Using analytic tools, the average generating functional can be represented as
\begin{align}\label{eq:zeta}
\mathbb{E}\left[W(j,h)\right]&=\sum_{k=1}^{\infty} \frac{(-1)^{k+1}c^k}{k k!}\,\mathbb{E}\left[Z^k(j,h)\right]  -\ln(c)+\gamma+R(c,j)
\end{align}
where the quantity $c$ is a dimensionless arbitrary constant, $\gamma$ is the Euler--Mascheroni constant, and $R(c)$ is given by
\begin{equation}
R(c,j)=-\int [dh]P(h)\int_{c}^{\infty}\,\dfrac{dt}{t}\, e^{-Z(j,h)t}.
\end{equation}
\noindent As can be checked directly from \ref{sec:appen}, after taking the functional derivatives, the contributions from $R(c,j)$ to the correlation functions can be neglected in the limit $c \to \infty$. Therefore, the series of partition-function moments generates both connected and disconnected correlation functions; see \ref{sec:appen}.  
%
%

We consider delta-correlated Gaussian disorder, i.e.,
$\mathbb{E}[h(x)h(y)]=\sigma^{2}\delta^{d}(x-y)$. To discuss the ordered phase of the model, i.e., the infrared regime, one expects the microscopic details of the disorder to be
irrelevant. After integrating over the disorder, each moment of the partition function,
$\mathbb{E}[Z^k(j,h)]$, can be written as
\begin{equation}
\mathbb{E}\left[Z^k(j,h)\right]=\int\,\prod_{i=1}^k\left[\d\varphi_{i}^{(k)}\right]\,\exp\left(-S_{\textrm{eff}}^{(k)}\left(\varphi_{i}^{(k)},j_{i}^{(k)}\right)\right),
\label{aa11}
\end{equation}
\noindent where the effective action $S_{\textrm{eff}}^{(k)}\left(\varphi_{i}^{(k)}\right)$ describes a field
theory with $k$ components. We note that, for fixed $k$, symmetry arguments and RG tools can be applied straightforwardly. However, extending RG methods and other QFT procedures to the sum over all $k$ requires care.
In what follows, we omit the superscript $(k)$ on the field variables; in
this notation, the effective action reads

\begin{align}\label{eq:Seff1}
S_{\textrm{eff}}^{(k)}(\varphi_{i},j_{i})= \int \d^{\,d}x &\left[\sum_{i=1}^{k}\left(\frac{1}{2}\,\varphi_{i}(x)\left(-\Delta+m_{0}^{2}\right)\varphi_{i}(x) +\frac{\lambda_{0}}{4!}(\varphi_{i}(x))^{4}\right) \right. \nonumber \\
&- \left.\frac{\sigma^{2}}{2}\sum_{i,j=1}^{k}\varphi_{i}(x)\varphi_{j}(x)-\sum_{i=1}^{k}\varphi_{i}(x)j_{i}(x)\right].
\end{align}
%

We note that the preceding series, together with the remainder, provides one possible representation of the DZF. This representation may fail when the series of moments grows faster than $e^{-kA}$, as in the random energy model, or when the moments cannot be expressed in closed form. Other representations can be obtained from the general definition in Eq.~(\ref{eq:zetager}). In what follows, we assume that the configuration considered is sufficiently well behaved to be represented by this series.

\section{Free theory of the model}

In view of Eq.~(\ref{eq:Seff1}), the propagator of our effective theory is a $k\times k$ matrix, i.e., it is non-diagonal. This feature has been explored in the literature in different
ways \cite{Lewandowski:2017omt,Lewandowski:2018bnn, Heymans:2023tgi, Heymans:2024dzq}. In the
context of the distributional zeta-function method, two approaches have been discussed.
First, one can adopt a diagonal ansatz in function space, $\varphi_i = \varphi_j$ for all $i,j$.
With this ansatz, perturbation theory can be carried out in the usual way, and consistent results
have been obtained for both the random-field and random-mass cases
\cite{Diaz:2016mto, Diaz:2017grg, Diaz:2017ilf, Soares:2019fed, Rodriguez-Camargo:2021ryf, Rodriguez-Camargo:2022wyz, Heymans:2022sdr}.
For instance, within the diagonal ansatz, one can recover the correct upper critical
dimension for the RFIM, $d_c^+=6$ \cite{binney1992theory}.

A natural question is what results can be obtained without assuming the diagonal ansatz. With
this in mind, a second approach was proposed \cite{Heymans:2023tgi, Heymans:2024dzq}: the diagonalization method. This approach
arises from analyzing the free part of the effective action
\begin{equation}
    \sum_{i,j=1}^kS_0(\varphi_i, \varphi_j) = \frac{1}{2} \sum_{i,j=1}^k\int \d^d x
    \varphi_i(x) \left(G^0_{ij} - \sigma^2\right)\varphi_j(x),
\end{equation}
where $G^0_{ij} =  \left(-\Delta + m_0^2\right)\delta_{ij}$. 
Such an action can be represented equivalently by
\begin{equation}\label{eq:free}
        \sum_{i,j=1}^k S_0 (\varphi_i, \varphi_j) = \frac{1}{2}\int \d^dx \,\, \inner{\Phi}{G\Phi},
\end{equation}
where $G$ is the $k\times k$ matrix whose components are $G^0_{ij}-\sigma^2$, $\Phi(x)$ is the vector whose components are $\varphi_i(x)$, and $\inner{\cdot}{\cdot}$ is the natural inner product on $\R^k$. Since $G$ is real symmetric, it can be diagonalized by an orthogonal matrix $O$:
\begin{equation}\label{eq:diag}
    D = \langle O, GO\rangle = \left[
\begin{array}{cccc}
  G^0_{11} -k\sigma^2 & 0& \cdots & 0 \\
   0 & G^0_{22} & \cdots & 0 \\
   \vdots & \cdots & \ddots & \vdots \\
   0 & \cdots & & G^0_{kk}
\end{array}
\right]_{k\times k}.
\end{equation}

First, we note that $\mathbb{R}^k$, which emerges from the disorder average, carries no special structure beyond that of a real vector space. In particular, Eq.~(\ref{eq:Seff1}) does not endow this space with any additional properties. Therefore, to keep the formulation as general as possible, we assume only the minimal structure on $\mathbb{R}^k$.
Defining $\tilde{\Phi}(x)=O\Phi(x)$ as the vector whose components are $\tilde{\Phi}=(\phi,\phi_1,\dots,\phi_{k-1})$, we can write a third form of the free effective action
\begin{align}
    &\sum_{i,j=1}^k S_0 (\varphi_i) = \frac{1}{2}\int \d^dx \,\, 
      \phi(x)(-\Delta + m_0^2 - k\sigma^2) \phi(x) \nonumber \\
      &\hspace{1.65cm} +   \frac{1}{2} \sum_{\alpha=1}^{k-1} \int \d^dx \,\, 
      \phi_\alpha(x)(-\Delta + m_0^2) \phi_\alpha(x),
\end{align}
which is clearly the sum of $k$ free actions with two distinct differential operators. Note that to distinguish the diagonal and non-diagonal bases, we changed the labels from $i,j$ to $\alpha, \beta$ in the equation above.

As we have seen, there is no difficulty in applying the diagonalization approach, Eq.~(\ref{eq:diag}), to the free effective action. The functional measure is also well-behaved under this change of variables: since the transformation matrix is orthogonal, the absolute value of the Jacobian is unity. The sources $j_i$, introduced to generate correlation functions, can likewise be chosen to transform according to the inverse transformation of the vector $\Phi$, so they remain well-behaved. In what follows, we consider the source-free case. Thus, for free actions, the diagonalization approach describes the system without any ansatz in functional space.
A problem arises once interactions are included.

It follows from Eq.~(\ref{eq:diag}) that there always exists a set of $k-1$ degenerate eigenvalues. This implies that the corresponding eigenvectors (the columns of $O$) must be orthogonalized. Using the notation $O_{i,j}$ for the $(i,j)$ entry of the matrix $O$, where $i$ and $j$ label the row and column, respectively, one can apply the Gram--Schmidt procedure to obtain the following set of eigenvectors
\begin{align}\label{eq:basis}
 O_{i,1} = \frac{1}{\sqrt{k}}, \quad O_{i,\alpha+1}=
\begin{cases}
\displaystyle\frac{1}{\sqrt{\alpha(\alpha+1)}}, & 1\le i\le\alpha,\\[6pt]
\displaystyle -\frac{\alpha}{\sqrt{\alpha(\alpha+1)}}, & i=\alpha+1,\\[6pt]
0, & i>\alpha+1,
\end{cases}
\end{align}
where $i \in \{1,2\dots, k\}$ and $\alpha \in \{1,2, \dots, k-1\}$. In general, for any representation, one can verify the orthogonality of the columns.

\section{Interacting theory}

As can be seen from Eq.~(\ref{eq:Seff1}), after the disorder average the effective interaction is not invariant under rotations in $\R^k$. This type of interaction is known in the literature as a cubic-anisotropic interaction \cite{aharony1973critical, Nattermann:1977uz, kleinert2001critical}. To write the interaction in the new basis, we first set
\begin{equation}
\varphi_i \;=\; \frac{\phi}{\sqrt{k}} + \sum_{\alpha=1}^{k-1} O_{i,\alpha+1}\,\phi_\alpha,
\end{equation}
for fixed $i$, we get
\begin{align}
\varphi_i^4 &= \left(\frac{\phi}{\sqrt{k}} + \sum_{\alpha} O_{i,\alpha+1}\phi_\alpha\right)^4 \\ \nonumber
&= \frac{\phi^4}{k^2}
+ 4\frac{\phi^3}{k^{3/2}} \sum_\alpha O_{i,\alpha+1}\phi_\alpha
+ 6\frac{\phi^2}{k}\sum_{\alpha,\beta} O_{i,\alpha+1}O_{i,\beta+1}\phi_\alpha\phi_\beta \\ \nonumber
&\quad + 4\frac{\phi}{\sqrt{k}} \sum_{\alpha,\beta,\gamma} O_{i,\alpha+1}O_{i,\beta+1}O_{i,\gamma+1}\,\phi_\alpha\phi_\beta\phi_\gamma \\\nonumber
&\quad + \sum_{\alpha,\beta,\gamma,\delta} O_{i,\alpha+1}O_{i,\beta+1}O_{i,\gamma+1}O_{i,\delta+1}\,
\phi_\alpha\phi_\beta\phi_\gamma\phi_\delta.
\end{align}

Summing over $i=1,\dots,k$, and using the orthogonality of the columns of $O_{i,\alpha+1}$, we obtain
\begin{align}
 \sum_{i=1}^k \varphi_i^4 = \frac{\phi^4}{k} + 6 \frac{\phi^2}{k}\sum_{\alpha = 1}^{k-1}\phi_\alpha^2 + \frac{3(k-1)}{k(k+1)}\sum_{\alpha, \beta =1}^{k-1}\phi_\alpha^2 \phi_\beta^2 + \sum_{\alpha, \beta, \gamma, \delta = 1}^kR_{\alpha \beta \gamma \delta}\phi_\alpha \phi_\beta \phi_\gamma \phi_\delta.
 \end{align}
Therefore, the effective action given in Eq.~(\ref{eq:Seff1}) can be rewritten as
\begin{align}\label{eq:acexact1}
    &S_{\mathrm{eff},1}^{(k)}(\phi,\phi_\alpha) =\int \d^dx \left[ \frac{1}{2} \phi(x)(-\Delta + m_0^2 - k\sigma^2) \phi(x) +  \frac{1}{2} \sum_{\alpha=1}^{k-1} \phi_\alpha(x)(-\Delta + m_0^2) \phi_\alpha(x)  \right. \nonumber \\
    &+ \left.\frac{1}{4!k} \left( \lambda\phi^4(x) + 6\lambda \phi^2(x) \sum_{\alpha=1}^{k-1}\phi_\alpha^2(x) + \lambda\frac{3(k-1)}{k+1}\sum_{\alpha, \beta =1}^{k-1}\phi_\alpha^2 \phi_\beta^2 + k\sum_{\alpha, \beta, \gamma, \delta = 1}^{k-1}R_{\alpha \beta \gamma \delta}\phi_\alpha \phi_\beta \phi_\gamma \phi_\delta\right)\right].
\end{align}
We note that the symmetry of this action is $\mathbb{Z}_2\times {\rm O}(k-1)$, with an anisotropic contribution from the last term, which can be viewed as a generalization of the usual cubic-anisotropic model considered in Refs.~\cite{Amit:1984ms, kleinert2001critical}. The first three interactions are the standard ones in the $\mathbb{Z}_2\times {\rm O}(k-1)$ model. The last term is a traceless anisotropic tensor that, due to the orthogonality of $O_{i\alpha}$ (see Eq.~(\ref{eq:basis})), can be specified in terms of the following independent components
\begin{align}
R_{\alpha\alpha\alpha\alpha}
&= \frac{1+\alpha^{3}}{\alpha(\alpha+1)^2}\lambda \;-\; \frac{3(k-1)}{k(k+1)}\lambda, \\[6pt]
R_{\alpha\alpha\beta\beta}
&= \frac{1}{\max(\alpha,\beta)\big(\max(\alpha,\beta)+1\big)}\lambda
\;-\; \frac{k-1}{k(k+1)}\lambda,
\qquad \alpha\neq\beta.
\end{align}
It is also interesting to note that for $k=1$ the theory is $\mathbb{Z}_2\times {\rm O}(k-1)$-invariant.

\section{RFIM Actions and Propagators}\label{sec:RFIM}

Let us recall that, for fixed $\sigma^2$, there exists a $k$ such that $k=k_c = \lfloor \frac{m^2}{\sigma^2}\rfloor$. Therefore, the series generated by the distributional zeta-function can be written as
\begin{equation}\label{eq:2log}
\mathbb{E}\left[\ln Z\right] = \sum_{k=1}^{k_c} c_{k}\mathbb{E}[Z_1^k] + \sum_{k=k_c+1}^\infty c_{k}\mathbb{E}[Z_2^k] + R(c,j),
\end{equation}
where $\mathbb{E}[Z_1^k]$ is associated with the effective action in Eq.~(\ref{eq:acexact1}) and $\mathbb{E}[Z_2^k]$ is associated with the action with shifted mass (note that in this case $m_0^2-k\sigma^2<0$). Using the results of Eqs.  (\ref{eq:restsecond}) and (\ref{eq:restthird}), the contribution of the correlation function of $R(c,j)$ can be neglected in the limit $c\to\infty$. Setting $\phi(x)=v-\psi(x)$, where $v^2=\frac{6k}{\lambda}(k\sigma^2-m_0^2)$, the shifted action can be obtained as
\begin{align}\label{eq:acexact2}
  S_{\mathrm{eff},2}^{(k)}(\phi,\phi_\alpha) &=\int \d^dx \left\{ \frac{1}{2} \psi(x)(-\Delta + m_1^2) \psi(x) +  \frac{1}{2} \sum_{\alpha=1}^{k-1} \phi_\alpha(x)(-\Delta + m_2^2) \phi_\alpha(x)   \right. \nonumber \\
  &+\left. \frac{\lambda}{4!} \left[ \frac{1}{k}\psi^4(x) + \frac{4v}{k}\psi^3(x) + \frac{12v}{k} \psi(x)\sum_{\alpha =1}^{k-1}\phi^2_\alpha + \frac{6}{k} \psi^2(x) \sum_{\alpha=1}^{k-1}\phi_\alpha^2(x) \right. \right. \nonumber \\ 
  &+ \left. \left. \frac{3(k-1)}{k+1}\sum_{\alpha, \beta =1}^{k-1}\phi_\alpha^2 \phi_\beta^2 + \sum_{\alpha, \beta, \gamma, \delta = 1}^{k-1}R_{\alpha \beta \gamma \delta}\phi_\alpha \phi_\beta \phi_\gamma \phi_\delta\right]\right\},
\end{align}
where $m_1^2 = 2(k\sigma^2 - m_0^2)$ and $m_2^2 = \frac{1}{2}(3k\sigma^2 - m_0^2)$. Therefore, in principle, the effective action of the RFIM always has two contributions: one associated with the symmetric phase (Eq.~(\ref{eq:acexact1})) and one associated with the symmetry-broken phase (Eq.~(\ref{eq:acexact2})).

Since the quenched average $\mathbb{E}\left[\ln Z\right]$, obtained by the distributional zeta-function method, does \emph{not} rely on an analytic-continuation procedure and the effective actions in Eqs.~(\ref{eq:acexact1})--(\ref{eq:acexact2}) are derived directly, all the properties of the action in Eq.~(\ref{eq:Seff1}) are preserved. Therefore, one expects that any procedure used to compute physical observables can be applied directly to Eq.~(\ref{eq:2log}).

A basic ingredient of different methods, especially perturbative ones, is the tree-level propagator. Explicit calculations show that the functional derivatives of Eq.~(\ref{eq:zeta}) with respect to external sources generate the connected correlation functions. The disconnected propagator can be obtained similarly, by differentiating with respect to different external sources. As can be explicitly computed, the contributions from $R(c,j)$ to the correlation functions go to zero as $e^{-c}$ as $c$ increases. Therefore, in the limit $c\to \infty$, the series in Eq.~(\ref{eq:2log}) is enough to reproduce the correlation functions of the model.

Since only one of the field variables carries explicit information about the disorder in the action in Eq.~(\ref{eq:acexact1}) (namely $\phi$), it is natural to first investigate its propagator and that of its counterpart in Eq.~(\ref{eq:acexact2}) (namely $\psi$). At tree level in momentum space, the propagators of $\phi$ (for $k\leq k_c$) and that of $\psi$ (for $k>k_c$) are as follows (see \ref{sec:appen}):
\begin{equation}
G^{(2)}(p) = \lim_{c\to\infty}G^{(2)}_0(p) = \lim_{c\to\infty}\left[ \sum_{k=1}^{k_c} \frac{c_k}{p^2 + m_0^2 - k\sigma^2} + \sum_{k= k_c+ 1}^{\infty} \frac{c_k}{p^2 + m_1^2}\right],
\end{equation}
where $c_k = (-1)^kc^{k}/ (k!k)$.
Let us introduce the following hypothesis: for any $k<k'$, we have $k\sigma^2\ll p^2+m^2$. We refer to this condition as the weak-disorder limit. Under this hypothesis, the $c$-dependent propagator can be rewritten as
\begin{equation}
G^{(2)}_0(p) = \sum_{k=1}^{k'} \frac{c_k}{p^2 + m_0^2 - k\sigma^2} + \sum_{k=k'+1}^{k_c} \frac{c_k}{p^2 + m_0^2 - k\sigma^2} + \sum_{k= k_c+ 1}^\infty \frac{c_k}{p^2 + m_1^2}.
\end{equation}
The first contribution to the propagator can be expressed as
\begin{align}\label{eq:propphi}
	\frac{1}{p^2 + m_0^2 - k\sigma^2} &= \frac{1}{p^2 + m_0^2}\frac{1}{1 - \frac{k\sigma^2}{p^2 + m_0^2}} = \sum_{n=0}^{\infty}\frac{(k\sigma^2)^{n}}{(p^2 + m_0^2)^{n+1}}\nonumber  \\
	&\Rightarrow \sum_{k=1}^{k_c} \frac{c_k}{p^2 + m_0^2 - k\sigma^2} = \sum_{n=0}^{\infty}\frac{(\sigma^2)^{n}}{(p^2 + m_0^2)^{n+1}} \sum_{k=1}^{k_c} \frac{(-1)^k k^{n-1}c^k}{ k!},
\end{align}
For fixed $c$, the coefficient in the $k$-series increases until $k\leq(n-1)/W(n-1)$, where $W(n-1)$ is the Lambert $W$ function, which is determined by the equation $(1+1/k)^{n-1}=k+1$. After that, it decreases super-exponentially.

We require $k\sigma^2$ to be small in order to justify the expansion. The $n=0$ contribution reproduces the propagator of the standard theory and therefore does not introduce anything new. The term with $n=1$ is the first contribution that modifies the infrared behavior of the model, yielding an upper critical dimension $d_c^+=6$. Around $d=6$, the remaining terms with $n>1$ are finite and irrelevant, and they are accompanied by higher powers of $k\sigma^2$.

However, this is not the propagator of our theory. For simplicity, let us assume that $k' = k_c$. Therefore, we can use $l = k - k_c$ to write
\begin{align}
 \sum_{k= k_c+ 1}^\infty \frac{c_k}{p^2 + m_1^2} = \sum_{l=1}^\infty \frac{c_l}{p^2 + 2(l\sigma^2 - m_0^2)} &<  \sum_{l=1}^\infty \frac{|c_l|}{p^2 +2(l\sigma^2 - m_0^2)} \nonumber \\ 
 &< \frac{1}{2(k_c+1)\sigma^2} \sum_{l=1}^{\infty} \frac{|c|^l}{l!l}.
\end{align}
Therefore, in the weak-disorder limit we have $k_c=\left\lfloor\frac{m_0^2}{\sigma^2}\right\rfloor\gg 1$, and these contributions converge rapidly to zero as $k_c$ increases (for each $c \in \mathbb{R}$). Thus, in this limit, they can be neglected, and we can take $k'=k_c\to\infty$. Consequently, taking only the contributions that survive in the $c\to \infty$ limit (see~\ref{sec:appen}), the propagator in the weak-disorder limit is given by
\begin{equation}\label{eq:2ptfunc}
G^{(2)}(p) =  \frac{1}{p^2 + m_0^2} + \frac{\sigma^2}{(p^2 + m_0^2)^2}.
\end{equation}

To analyze the disorder contribution to this propagator, one could reconstruct an effective action in which disorder enters as a vertex and compute the corresponding beta function for the effective disorder coupling. Alternatively, this propagator can be used to develop perturbation theory in powers of the coupling $\lambda$ in the effective action, Eq.~(\ref{eq:acexact1}). The important point is that, in the weak-disorder limit, the upper critical dimension of the theory is $6$, in agreement with the literature (see, for example, Ref.~\cite{Rychkov:2023rgq}).

In the strong-disorder limit, defined by $\sigma^2 \approx m_0^2$, the critical index
$k_c = \left\lfloor \frac{m_0^2}{\sigma^2} \right\rfloor$ is reduced to $k_c = 1$.
Consequently, the lowest contribution in the distributional zeta-function expansion already
corresponds to a finite disorder-induced mass scale, indicating that the system does not
approach criticality through a gradual softening of modes. Instead, the spectrum generated by
quenched disorder places the system directly in a gapped regime, a mechanism naturally captured
by the distributional zeta-function method. The second contribution to the series in Eq. (\ref{eq:2log}), corresponding to $k>k_c$, is always present in any real system. The limiting case, $\sigma^2 \approx m_0^2$, therefore restricts the domain of validity of the present analysis.

Before analyzing the strong-disorder case, let us discuss the physical meaning of our definitions. As seen from Eq.~(\ref{eq:2log}), the quenched free energy is composed of contributions associated with the weak- and strong-disorder regimes. The weak-disorder contribution corresponds to the action in Eq.~(\ref{eq:acexact1}), and the corresponding ground states are located at the origin ($m_0^2 - k\sigma^2 \geq 0$). That is, the first $k_c$ terms of the series representation have a common ground state. By contrast, the remaining terms of the series (the strong-disorder regime) correspond to the action defined in Eq.~(\ref{eq:acexact2}). The corresponding free energy has distinct ground states that, for each $k>k_c$, are shifted away from the origin. This collection of distinct ground states defines the quenched free-energy landscape.

Within these definitions, the weak- and strong-disorder limits correspond to whether or not the ground state of the system is localized at the origin. The weak-disorder limit is suitable for systems in which the ground state can be taken to be at the origin, while the strong-disorder limit must be applied whenever the metastable state cannot be located at the origin.

In the strong-disorder limit, the free two-point function is given by
\begin{equation}
G^{(2)}_0(p)
= \sum_{k=1}^{\infty} \frac{c_k}{p^2 + m_k^2}
= \sum_{k=1}^{\infty} \frac{c_k}{p^2 + 2(k\sigma^2 - m_0^2)},
\end{equation}
where all effective masses increase linearly with $k$. The resulting spectral structure, together with the statistical suppression encoded in the coefficients $c_k$, ensures rapid convergence of the series and a well-defined infrared limit for the tree-level propagator. Since, at tree-level, no massless pole develops, there is no need to reorganize the theory by means of an infrared expansion or to modify its ultraviolet behavior, which remains that of a scalar field theory with upper critical dimension $d_c=4$.

In the case of symmetry breaking, these features indicate that strong quenched disorder effects qualitatively modify the nature of the phase transition at tree-level. Rather than exhibiting a genuine second-order transition, characterized by a diverging correlation length and scale invariance, the system evolves smoothly between phases. Although the correlation length increases as the disorder strength is enhanced, it remains finite, reflecting the absence of a critical fixed point in the tree-level approximation. In this sense, strong disorder effects ``smear out'' the transition at the tree-level, replacing critical behavior with a disorder-driven crossover---a phenomenon closely related to Griffiths-type effects in disordered systems \cite{gri1,vojta2}. While the scenario discussed above is valid for the tree-level spectrum result, whether a critical fixed point exists remains an open question and a more robust analysis requires an RG treatment. This is left for future work.

Although the previous discussion rules out the possibility of a second-order phase transition from a symmetric to a broken-symmetry phase, going beyond the tree-level regime may allow symmetry restoration through a second-order phase transition. In such a case, the vacuum is shifted back to the origin, leading to the weak-disorder limit. A careful analysis of symmetry restoration requires the application of the RG to the interacting theory in the strong-disorder limit.

\section{Renormalization Group scaling properties of weak disorder effect limit }

Let us analyze the scaling properties of the correlation function in the infrared theory ($m_0 \to 0$) in the weak-disorder limit. That is, we consider the contributions to the quenched free-energy landscape for which all ground states are located at the origin. In this regime, one sees from Eq.~(\ref{eq:2ptfunc}) that the correlator $\langle \phi \phi\rangle$ behaves as a correlated composition of two fields with distinct scaling properties, i.e., a field decomposition with fixed scaling laws; that is, under a scaling transformation $\mu$, we get	
\begin{equation}
    \phi(x) \to \mu^{-(d/2 -2)}X(\mu x) + \mu^{-d/2}Y(\mu x).
\end{equation}

To obtain such a result, we propose a decomposition of $\phi$ into two components, $\phi = X + Y$. To understand this decomposition, we note that Eq.~(\ref{eq:2ptfunc}) has two distinct scaling laws; therefore, it cannot be represented by a single field with a fixed scaling law. The proposed decomposition is therefore introduced to organize the scaling of the relevant fields of the theory. The correlation function is then
$\langle \phi \phi\rangle = \langle XX \rangle + 2\langle XY \rangle + \langle YY \rangle$,
which should match the right-hand side of Eq.~(\ref{eq:2ptfunc}). If we choose
$\langle XX \rangle \propto p^{-2}$, $\langle YY \rangle \propto p^{-4}$, and $\langle XY \rangle = 0$,
this would be inconsistent with Eq.~(\ref{eq:2ptfunc}), since it would imply that the field
$\phi$ consists of two uncorrelated fields. Therefore, one must have
$\langle XY\rangle \neq 0$.

Now let us analyze the scaling dimensions of the correlations. Denote by $[A]$ the scaling
dimension of the quantity $A$. Using the general scaling form of a correlation function in
the IR, we obtain
\begin{equation}
  \langle XX \rangle \propto p^{2[X] -d} \Rightarrow 2[X] -d = -2, \,\, \mathrm{or }, \, -4.
\end{equation}
If $2[X] -d = -2$, then we would need $\langle XY\rangle \propto p^{-4}$, and the correlation between the auxiliary fields $X$ and $Y$ would be more singular than the Gaussian part. This would lead to an unstable theory in these variables. Therefore, for consistency, we must take $2[X] -d = -4$, which is equivalent to $\langle XX \rangle \propto p^{-4}$. Consequently, we obtain that $\langle XY \rangle \propto p^{-2}$. We could also have $\langle YY \rangle \propto p^{-2}$ (note that any other power is not allowed by Eq.~(\ref{eq:2ptfunc})), but this can be consistently incorporated into $\langle XY \rangle$ by a change of variables. To see this explicitly, one can perform the following change of basis, $Y \to Y' = Y + \alpha p^2 X$, and compute $\langle Y'Y'\rangle$. In the explicit computation, one notices that $\alpha$ can always be chosen to ensure that the coefficient of $\langle Y'Y'\rangle$ vanishes. Therefore, without loss of generality, we set $\langle YY \rangle =0$. Thus, we have that
\begin{equation}
    \langle XX\rangle \propto \frac{1}{p^4}, \qquad \langle XY \rangle \propto \frac{1}{p^2}, \quad \langle \phi_\alpha \phi_\beta\rangle \propto \frac{1}{p^2},
\end{equation}
where the two-point function $\langle \phi_\alpha \phi_\beta\rangle$ is obtained directly from the effective action in Eq.~(\ref{eq:acexact1}). Thus, we have been able to separate the fields appearing in the effective action into different components with well-behaved scaling properties in the IR limit. A similar decomposition was first proposed by Cardy \cite{CARDY1985123}.

We can now rewrite the effective action in Eq.~(\ref{eq:acexact1}) in terms of the fields $X$, $Y$, and $\phi_\alpha$. Since $\langle YY \rangle$ can always be set to zero---that is, we may omit the term $Y(-\Delta)Y$---we can write the Gaussian part of the effective action as
\begin{equation}
    S_{\mathrm{eff},1-0}^{(k)-\mathrm{IR}}(X,Y, \phi_\alpha) = \int \d ^d x \left[ Y( -\Delta)X - \frac{k \sigma^2}{2}X^2 + \frac{1}{2}\sum_{\alpha=1}^{k-1}\phi_\alpha(-\Delta)\phi_\alpha\right].
\end{equation}
The interaction can be treated term by term. Let us first analyze the interaction proportional to $R_{\alpha\beta\gamma\delta}$. Near the upper critical dimension $d_c^+=6$, we can use the scaling of $\phi_\alpha$, $[\phi_\alpha]=d/2-1$, to obtain $[R_{\alpha\beta\gamma\delta}] = 4 - (6 - \epsilon) = -2+ \epsilon <0$,
which makes this contribution irrelevant. Therefore, the last term in Eq.~(\ref{eq:acexact1}) can be neglected in the IR limit. The same can be verified explicitly for the interaction $\sum_{\alpha,\beta} \phi_\alpha^2 \phi_\beta^2$, showing that $[\lambda] < 0$ near $d=6$, and therefore that this interaction is also irrelevant.

We can now expand the quartic interaction and use the scaling of $X$ and $Y$ to obtain the following behavior:
\begin{align}\label{eq:intdec}
\lambda\phi^4 = \lambda(X+Y)^4 &= \lambda X^4 + 4\lambda X^3Y + 6\lambda X^2Y^2 + 4\lambda XY^3 + \lambda Y^4 \nonumber \\
&=\lambda_X X^4 + \lambda_{XY}X^3Y + \lambda_{XY}^{(2)}X^2Y^2 + \lambda_{XY}^{(3)}XY^3 + \lambda_Y Y^4,
\end{align}
A similar analysis can be performed for the coupling $6\lambda \phi^2 \sum_\alpha\phi_\alpha^2$.
\setlength{\arrayrulewidth}{0.8pt}

\begin{table}[h]
\centering
\begin{tabular}{c c c}
\hline
Coupling & Scaling Dimension at $d_c^+=6$ & RG Status \\
\hline
$\lambda_X$ & $+2$ & Relevant \\
$\lambda_{XY}$ & $0$ & Marginal \\
$\lambda_{XY}^{(2)}$ & $-2$ & Irrelevant \\
$\lambda_{XY}^{(3)}$ & $-4$ & Irrelevant \\
$\lambda_Y$ & $-6$ & Irrelevant \\
\hline
\end{tabular}
\caption{Relevancy of interaction couplings near the weak-disorder upper critical dimension $d_c^+=6$.}
\end{table}

\setlength{\arrayrulewidth}{0.8pt}

\begin{table}[h]
\centering
\begin{tabular}{c c c}
\hline
Coupling & Scaling Dimension at $d_c^+=6$ & RG Status \\
\hline
$\lambda_{X^2}$ & $-2$ & Irrelevant \\
$\lambda_{X^2Y}$ & $-4$ & Irrelevant \\
$\lambda_{Y^2}$ & $-6$ & Irrelevant \\
\hline
\end{tabular}
\caption{Relevancy of couplings generated from $6\lambda\,\phi^2\sum_{\alpha}\phi_\alpha^2$ at $d_c^+=6$.}
\end{table}

\setlength{\arrayrulewidth}{0.4pt}

Therefore, we can now write the IR-relevant effective action of the RFIM as
\begin{align}\label{eq:iract}
S_{1}^{\mathrm{IR}}(X,Y, \phi_\alpha) = \int \d ^d x \Bigg[ &Y(x)( -\Delta)X(x) - \frac{k \sigma^2}{2}X(x)^2 + \frac{1}{2}\sum_{\alpha=1}^{k-1}\phi_\alpha(x)(-\Delta)\phi_\alpha(x) \nonumber \\
& + \lambda_X X^4(x) + \lambda_{XY}^{(1)} X^3(x) Y(x)\Bigg].
\end{align}

Furthermore, we note that once the microscopic theory gives rise to an interaction $\lambda>0$, analyzing Eq.~(\ref{eq:intdec}) implies a constraint on the magnitudes of the interactions. Denote the dimensionless magnitude of a quantity $A$ by $\tilde{A}$. Then, we must have
\begin{equation}\label{eq:intcons}
    \tilde{\lambda} = \tilde{\lambda}_X, \quad \mathrm{and}\,\,\,\,\, \tilde{\lambda}_X = 4\tilde{\lambda}_{XY}
\end{equation}
as an additional constraint. This constraint is closely related to the supersymmetric approach to the RFIM. Within Parisi--Sourlas supersymmetry, the quartic interactions can be systematically decomposed into operators classified as ``supersymmetry-writable'', ``supersymmetry-null'', and ``non-supersymmetry-writable''~\cite{Kaviraj:2020pwv}. The unique supersymmetry-writable quartic interaction is given, up to an overall normalization, by the irreducible combination
$\varphi^4 + 4\,\varphi^3 \omega + 6\,\varphi^2 \omega^2 + 4\,\varphi \omega^3 + \omega^4$,
in the notation of Ref.~\cite{Kaviraj:2020pwv}.
Thus, our IR effective action naturally encodes this constraint on the magnitudes of the interactions.

Our previous results show that the DZF method reproduces the behavior of the RFIM in the weak-disorder limit within a single formalism. Since approximations are introduced only after deriving the effective action in Eq.~(\ref{eq:acexact1}), we have shown that the action in Eq.~(\ref{eq:acexact1}), together with the constraint in Eq.~(\ref{eq:intcons}), encodes information about the infrared (IR) regime of the theory in the weak-disorder limit. Furthermore, the previous analyses provide a starting point for identifying the relevant operators of the RFIM in any dimension, especially in $d=3$.

This set of results also strengthens our tree-level conclusions concerning symmetry breaking in the strong-disorder regime of the RFIM. Since the strong-disorder action in Eq.~(\ref{eq:acexact2}) is as rigorous as its weak-disorder counterpart, it likewise encodes the critical behavior of the model within its domain of validity. In the symmetry-breaking case, the system undergoes the transition smoothly, without developing singularities in the two-point function. However, this does not rule out the possibility that disorder-driven symmetry restoration may still correspond to a genuine second-order phase transition.

Finally, we emphasize that obtaining the effective actions in both regimes enables not only the qualitative analyses presented here, but also the direct application of standard quantum field--theoretic methods to the RFIM.

\section{Conclusions}

In this work, we have developed a formulation of the continuous random-field Ising model based on the distributional zeta-function method. We have shown that it provides a new description of both weak- and strong-disorder regimes. In the weak-disorder limit, we derived the disorder-averaged effective action and demonstrated that its infrared structure dynamically generates the correlated scaling fields responsible for the characteristic $1/p^{4}$ behavior, the shift of the upper critical dimension to $d_c^{+}=6$, and the Cardy decomposition. This allowed us to construct a minimal infrared effective theory that encodes the universal critical behavior of the RFIM and clarifies the origin of its relevant operators and couplings.

In the strong-disorder regime, we obtained a diagonal quadratic action with a discrete spectrum of massive modes, showing that, at the symmetry-breaking transition, disorder averaging suppresses conventional criticality by eliminating massless excitations in tree-level approximation while still allowing for a smooth disorder-driven crossover. The resulting spectral representation of correlation functions converges rapidly and provides analytical control over the infrared behavior of the model in a regime that is usually inaccessible to perturbative approaches.

As we have shown, the weak-disorder effective action recovers the infrared action that dominates the critical physics near the usual upper critical dimension of the RFIM. We started with the standard continuous RFIM action and derived the weak-disorder action in closed form, which, to the best of our knowledge, is new in the literature. This action has a well-defined ground state located at the origin. Moreover, we extended the analysis of the free-energy landscape by also obtaining a closed effective action in the strong-disorder regime. We have also shown that metastable ground states far from the origin have the upper critical dimension given by $d_c^+ =4$. The analysis of such metastable ground states requires further investigation. We also note that, in the regime $\sigma^2 \gg m_0^2$, the series representation is no longer appropriate. An analysis of this regime requires a more suitable representation based on Eq.~(\ref{eq:zetager}).
Taken together, these results show that the distributional zeta-function method provides a framework that exposes the infrared degrees of freedom and mechanisms governing phase transitions in disordered systems in the tree-level scenario. By unifying weak- and strong-disorder physics within a single formalism and yielding the effective actions prior to approximation, our approach opens the door to systematic renormalization-group calculations.

\section*{Acknowledgments} 
The authors are grateful to S. A. Dias and G. Krein for fruitful discussions. This work was partially supported by Conselho Nacional de Desenvolvimento Cient\'{\i}fico e Tecnol\'{o}gico (CNPq), grants nos. 305000/2023-3 (N.F.S.), 311300/2020-0 (B.F.S.), and 307626/2022-9 (A.M.S.M.), and Fundação Carlos Chagas Filho de Amparo à Pesquisa do Estado do Rio de Janeiro (FAPERJ) grant no. E-26/203.318/2017 (B.F.S.). G.O.H. thanks Fundação Carlos Chagas Filho de Amparo à Pesquisa do Estado do Rio de Janeiro (FAPERJ) for the Ph.D. scholarship.  

\appendix

\section{Distributional zeta-function derivation of the free propagators}\label{sec:appen}

In the main text, we obtained the free propagators from the effective replicated action. The purpose of this appendix is to show how the same expressions follow directly from the distributional zeta-function (DZF) representation. This derivation also clarifies the role of the integer moments entering the DZF expansion and explains why the physical quenched propagators are not obtained from a direct sum of fixed-$k$ propagators.

\subsection{Connected propagator}

The disorder-averaged connected generating functional is
\begin{equation}
\mathbb{E}\!\left[W(j,h)\right]
=
\mathbb{E}\!\left[\ln Z(j,h)\right].
\end{equation}
Using the DZF representation, one has
\begin{equation}
\mathbb{E}\!\left[W(j,h)\right]
=
\sum_{k=1}^{\infty}
\frac{(-1)^{k+1}c^k}{k!\,k}
\,\mathbb{E}\!\left[Z(j,h)^k\right]
-\ln c-\gamma
+R(c,j),
\label{eq:DZFappendix}
\end{equation}
where
\begin{equation}
R(c,j)
=
-
\mathbb{E}\!\left[
\int_c^\infty
\frac{dt}{t}
e^{-tZ(j,h)}
\right].
\end{equation}
The essential point is that Eq.~(\ref{eq:DZFappendix}) is an identity for the logarithm. Therefore, the quantity entering the DZF expansion is the unnormalized moment
\begin{equation}
\mathbb{E}\!\left[Z(j,h)^k\right],
\end{equation}
rather than the normalized propagator associated with the $k$-th replicated theory.
For the Gaussian RFIM,
\begin{equation}
\mathbb{E}\!\left[Z(j,h)^k\right]
=
\mathcal Z_k[j],
\end{equation}
where $\mathcal Z_k[j]$ is generated by the quadratic effective action discussed in Sec.~\ref{sec:RFIM}. Choosing the same source in every replica,
\begin{equation}
j_1=\cdots=j_k=j,
\end{equation}
one finds
\begin{equation}
\mathcal Z_k[j]
=
\mathcal Z_k[0]
\exp\left\{
\frac12
\int\frac{d^dp}{(2\pi)^d}
\,j(p)
\frac{k}{p^2+m_0^2-k\sigma^2}
j(-p)
\right\}.
\label{eq:Zkappendix}
\end{equation}
Taking two functional derivatives gives
\begin{equation}
\left.
\frac{\delta^2\mathcal Z_k[j]}
{\delta j(p)\delta j(-p)}
\right|_{j=0}
=
\mathcal Z_k[0]
\frac{k}
{p^2+m_0^2-k\sigma^2}.
\label{eq:seriespiece}
\end{equation}
Substituting Eq.~(\ref{eq:seriespiece}) into Eq.~(\ref{eq:DZFappendix}) yields the series contribution
\begin{equation}
G_{\mathrm{conn}}^{(\mathrm{series})}(p)
=
\sum_{k=1}^{\infty}
\frac{(-1)^{k+1}c^k}{k!}
\,
\frac{\mathcal Z_k[0]}
{p^2+m_0^2-k\sigma^2}.
\label{eq:seriesconn}
\end{equation}
Equation~(\ref{eq:seriesconn}) is not the physical connected propagator. The reason is that differentiating the series produces terms proportional to both the first and second functional derivatives of $W(j,h)$. These extra contributions are canceled by the functional derivatives of the remainder $R(c,j)$.
To see this explicitly, write the DZF identity for a positive variable $z$,
\begin{equation}
\ln z
=
\sum_{k=1}^{\infty}
\frac{(-1)^{k+1}c^k}{k!\,k}
z^k
-\ln c-\gamma
+r_c(z),
\end{equation}
where
\begin{equation}
r_c(z)
=
-
\int_c^\infty
\frac{dt}{t}
e^{-tz}.
\end{equation}
Differentiating once gives
\begin{equation}
\frac{d}{dz}
\left[
\sum_{k=1}^{\infty}
\frac{(-1)^{k+1}c^k}{k!\,k}
z^k
\right]
=
\frac{1-e^{-cz}}{z},
\end{equation}
while
\begin{equation}
\frac{dr_c}{dz}
=
\frac{e^{-cz}}{z}.
\end{equation}
Therefore,
\begin{equation}
\frac{d}{dz}
\left[
\sum_{k=1}^{\infty}
\frac{(-1)^{k+1}c^k}{k!\,k}
z^k
+r_cz)
\right]
=
\frac{1}{z},
\end{equation}
as required.
Applying the chain rule with $z=Z(j,h)$, one obtains
\begin{align}
\frac{\delta^2}{\delta j(x)\delta j(y)}
\sum_{k=1}^{\infty}
\frac{(-1)^{k+1}c^k}{k!\,k}
Z(j,h)^k
=
&(1-e^{-cZ_h})
W_{xy}
\nonumber\\
&
-
c\,e^{-cZ_h}
W_xW_y,
\label{eq:seriessecond}
\end{align}
where
\[
W_x=\frac{\delta W(j,h)}{\delta j(x)},
\qquad
W_{xy}
=
\frac{\delta^2W(j,h)}
{\delta j(x)\delta j(y)}.
\]
The remainder contributes
\begin{align}\label{eq:restsecond}
\frac{\delta^2r_c(Z_h)}
{\delta j(x)\delta j(y)}
=
&e^{-cZ_h}
W_{xy}
+c\,e^{-cZ_h}
W_xW_y.
\end{align}
Adding Eqs.~(\ref{eq:seriessecond}) and (\ref{eq:restsecond}) immediately gives
\begin{equation}
\frac{\delta^2}
{\delta j(x)\delta j(y)}
\left[
\sum_{k=1}^{\infty}
\frac{(-1)^{k+1}c^k}{k!\,k}
Z(j,h)^k
+r_c(Z_h)
\right]
=
W_{xy},
\end{equation}
which proves that the DZF reconstructs the connected generating functional after the remainder is included.
For the free theory,
\begin{equation}
W_{xy}(p)
=
\frac{1}{p^2+m_0^2},
\end{equation}
and therefore
\begin{equation}
G_{\mathrm{conn}}^{(0)}(p)
=
\frac{1}{p^2+m_0^2},
\end{equation}
in agreement with the result obtained in the main text.

\subsection{Disconnected propagator}

The disconnected propagator is defined by the disorder average of the product of generating functionals,
\begin{equation}
\mathbb{E}\!\left[W(j,h)W(j',h)\right].
\end{equation}
Introducing two independent external sources, the DZF representation becomes
\begin{align}
\mathbb{E}\!\left[W(j,h)W(j',h)\right]
=&
\sum_{k,\ell=1}^{\infty}
\frac{(-1)^{k+\ell}c^{k+\ell}}
{k!\,\ell!\,k\,\ell}
\mathbb{E}\!\left[
Z(j,h)^k
Z(j',h)^\ell
\right]
\nonumber\\
&
+\mathcal{R}_c^{(2)}[j,j'],
\label{eq:DZF2}
\end{align}
where $\mathcal{R}_c^{(2)}$ denotes the corresponding remainder.
The relevant quantity is therefore the mixed integer moment
\begin{equation}
\mathcal{Z}_{k,\ell}[j,j']
=
\mathbb{E}\!\left[
Z(j,h)^k
Z(j',h)^\ell
\right].
\end{equation}
For the Gaussian random-field model, integrating over the disorder gives
\begin{align}
\mathcal{Z}_{k,\ell}[j,j']
=
\mathcal{Z}_{k,\ell}[0]
\exp\Bigg\{
&
\frac12
(j,\mathcal{G}_k j)
+
\frac12
(j',\mathcal{G}_\ell j')
\nonumber\\
&
+
(j,\mathcal{G}_{k\ell}j')
\Bigg\},
\label{eq:Zkl}
\end{align}
where the mixed kernel is
\begin{equation}
\mathcal{G}_{k\ell}(p)
=
\frac{k\ell\,\sigma^2}
{\left(p^2+m_0^2-k\sigma^2\right)
\left(p^2+m_0^2-\ell\sigma^2\right)}.
\label{eq:mixedkernel}
\end{equation}
Taking one functional derivative with respect to each source yields
\begin{equation}
\left.
\frac{\delta^2
\mathcal{Z}_{k,\ell}}
{\delta j(p)\delta j'(-p)}
\right|_{j=j'=0}
=
\mathcal{Z}_{k,\ell}[0]
\,
\mathcal{G}_{k\ell}(p).
\label{eq:mixedderivative}
\end{equation}
Substituting Eq.~(\ref{eq:mixedderivative}) into Eq.~(\ref{eq:DZF2}) gives the series contribution
\begin{align}
G_{\mathrm{disc}}^{(\mathrm{series})}(p)
=
\sum_{k,\ell=1}^{\infty}
\frac{(-1)^{k+\ell}c^{k+\ell}}
{k!\,\ell!}
\,
\frac{\mathcal{Z}_{k,\ell}[0]\sigma^2}
{\left(p^2+m_0^2-k\sigma^2\right)
\left(p^2+m_0^2-\ell\sigma^2\right)}.
\label{eq:discseries}
\end{align}
As in the connected case, Eq.~(\ref{eq:discseries}) is not the physical disconnected propagator. The functional derivatives of the series generate additional terms involving the first derivatives of the generating functional, which are canceled identically by the corresponding derivatives of the remainder $\mathcal{R}_a^{(2)}$.
Indeed, differentiating the DZF identity twice with respect to two independent sources gives
\begin{align}
\frac{\delta^2}
{\delta j(x)\delta j'(y)}
\left[
\sum_{k,\ell}
\frac{(-1)^{k+\ell}c^{k+\ell}}
{k!\,\ell!\,k\,\ell}
Z(j,h)^k
Z(j',h)^\ell
\right]
=
&
\left(1-e^{-cZ_h}\right)
W_xW'_y
\nonumber\\
&
-
c\,e^{-cZ_h}
W_xW'_y,
\end{align}
where
\[
W_x=\frac{\delta W(j,h)}{\delta j(x)},
\qquad
W'_y=\frac{\delta W(j',h)}{\delta j'(y)}.
\]
The corresponding contribution from the remainder is
\begin{align}\label{eq:restthird}
\frac{\delta^2\mathcal{R}_c^{(2)}}
{\delta j(x)\delta j'(y)}
=
e^{-cZ_h}
W_xW'_y
+
c\,e^{-cZ_h}
W_xW'_y,
\end{align}
so that the exponential terms cancel. Consequently,
\begin{equation}
\frac{\delta^2}
{\delta j(x)\delta j'(y)}
\mathbb{E}\!\left[
W(j,h)W(j',h)
\right]
=
\mathbb{E}
\!\left[
W_xW'_y
\right],
\end{equation}
which is the definition of the disconnected two-point correlation function.
For the Gaussian theory considered in the main text, the resulting tree-level disconnected propagator is
\begin{equation}
G_{\mathrm{disc}}^{(0)}(p)
=
\frac{\sigma^2}
{\left(p^2+m_0^2\right)^2},
\end{equation}
in agreement with the expression obtained from the replicated effective action.
Combining the connected and disconnected contributions finally gives
\begin{equation}
G^{(0)}(p)
=
G_{\mathrm{conn}}^{(0)}(p)
+
G_{\mathrm{disc}}^{(0)}(p)
=
\frac{1}{p^2+m_0^2}
+
\frac{\sigma^2}
{\left(p^2+m_0^2\right)^2}.
\end{equation}
As we can observe from Eqs. (\ref{eq:restsecond}) and (\ref{eq:restthird}), the contributions of the remainder are proportional to $e^{-cZ}$, therefore in the limit $c \to \infty$ only the series contributes and the Eq. (\ref{eq:2ptfunc}) follows immediately.

\printbibliography

\end{document}